\documentclass[pre, superscriptaddress, twocolumn, floatfix]{revtex4}
\usepackage{hyperref}
\usepackage{calc,graphicx,subfigure}
\usepackage[usenames]{color}
\definecolor{Red}{rgb}{1,0,0}

\newcommand{\LASSP}{Laboratory of Atomic and Solid State Physics, Cornell University}
\newcommand{\UCD}{UCD Conway Institute of Biomolecular \& Biomedical Research, University College Dublin}
\newcommand{\CTC}{Cornell Theory Center, Cornell University}
\newcommand{\MBG}{Department of Molecular Biology and Genetics, Cornell University}

\begin{document}

\title{Extracting falsifiable predictions from sloppy models}
\author{Ryan N. Gutenkunst}
\affiliation{\LASSP}
\author{Fergal P. Casey}
\affiliation{\UCD}
\author{Joshua J. Waterfall}
\affiliation{\MBG}
\author{Christopher R. Myers}
\affiliation{\CTC}
\author{James P. Sethna}
\affiliation{\LASSP}

\begin{abstract}
Successful predictions are among the most compelling validations of any model. Extracting falsifiable predictions from nonlinear multiparameter models is complicated by the fact that such models are commonly \emph{sloppy}, possessing sensitivities to different parameter combinations that range over many decades. Here we discuss how sloppiness affects the sorts of data that best constrain model predictions, makes linear uncertainty approximations dangerous, and introduces computational difficulties in Monte-Carlo uncertainty analysis. We also present a useful test problem and suggest refinements to the standards by which models are communicated. 
\end{abstract}

\maketitle

Reverse engineering of biological networks entails working from data to models, and successful predictions are among the most important and compelling validations of those models.
Making useful predictions, however, entails working back from models to data.
A model's behavior depends both on its structure (components and interactions) and on its parameters (numbers quantifying the structure).
In biology the focus is on structure; parameter values themselves are generally of little interest.
To test a model structure, quantitative predictions must be tempered by rigorous estimates of their uncertainties, accounting for model behavior over all sets of parameters consistent with the available data~\cite{Brown2003a}.
An experimental result inconsistent with these uncertainties is then strong evidence that some assumption in the model structure is false.
We argue that any valuable assessment of reverse engineering methods in systems biology needs to address protocols and algorithms for evaluating model prediction uncertainty. 

Here we detail how to extract falsifiable predictions from complex biological models. 
We focus on complications introduced by \emph{sloppiness}, the presence of orders of magnitude variation in sensitivity to different parameter combinations~\cite{Brown2003a}.
Complex biological models appear to be universally sloppy~\cite{Gutenkunst2006}, along with many other nonlinear multiparameter models~\cite{Waterfall2006}.
Sloppiness affects the sorts of data that best constrain model predictions~\cite{Gutenkunst2006}, makes linear uncertainty approximations dangerous, and introduces computational difficulties in Monte-Carlo uncertainty analysis.
We discuss all these issues, introduce a useful test system, and suggest refinements to the standards by which models are communicated.

\section{Sloppiness}

\begin{figure}
\begin{center}
\includegraphics{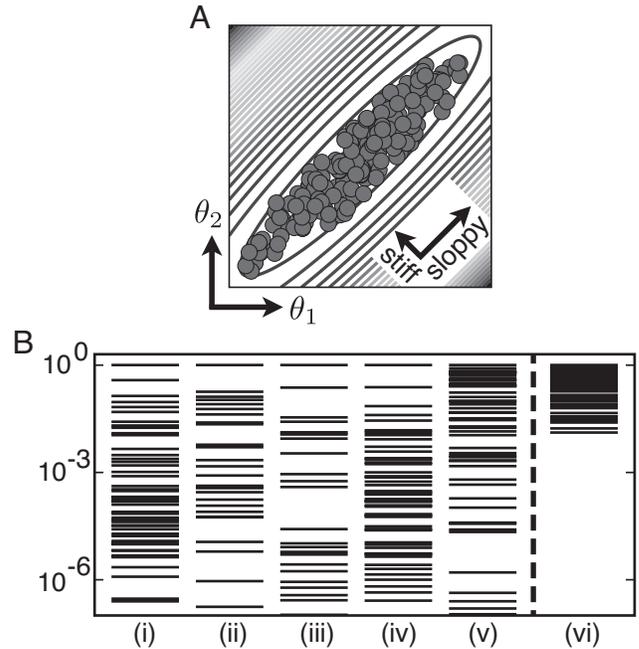}\label{fig:banana}
\caption{
Conceptual illustration and quantification of sloppiness. 
A) Contours measure the change in model behavior as parameters vary.
The model is sensitive to \emph{stiff} directions and insensitive to \emph{sloppy} directions.
Dots indicate a region of parameter sets consistent with available data; the region is naturally aligned along sloppy directions, so predictions from this data can have small uncertainty, even though parameter uncertainty is large.
B) Sloppy eigenvalue spectra for models of: (i) growth factor signaling~\cite{Brown2004}, (ii) intra-receptor dynamics~\cite{Edelstein1996}, (iii) circadian rhythms~\cite{Leloup1999}, (iv) quantum Monte-Carlo wave functions, (v) sums of exponentials.
Fitting a plane to data (vi), on the other hand, is not a sloppy problem.
}
\end{center}
\end{figure}

Sloppiness is illustrated conceptually in figure~\ref{fig:banana}A, which shows a plot in parameter space where contours represent surfaces of constant model behavior.
The model is very sensitive to \emph{stiff} parameter combinations and very insensitive to \emph{sloppy} combinations.

These sensitivities can be quantified by a \emph{cost} function $C\left(\theta\right)$ that measures the change in a model's behavior as parameters $\theta$ vary.
We Taylor expand $C\left(\theta\right)$ about a set of best-fit parameters $\theta^*$ yielding the Hessian matrix:
\begin{equation}
H_{ij}\left(\theta^*\right) = \left.\frac{\partial^2 C}{\partial \theta_i \partial \theta_j} \right|_{\theta^*}.\label{eqn:hess}
\end{equation}
The eigenvectors of $H$ are the principle axes of the ellipses shown in figure~\ref{fig:banana}A, and the model's sensitivity to parameter changes along each axis is proportional to the square root of the corresponding eigenvalue.

Figure~\ref{fig:banana}B shows eigenvalue spectra for several models (a subset
of those in~\cite{Gutenkunst2006} and~\cite{Waterfall2006}).
Column (i) is a model of growth factor signaling in PC12 cells~\cite{Brown2004}, where the cost function measures the model deviation from a set of real data (48 parameters and 68 data points).
Columns (ii) and (iii) are models of intra-receptor dynamics~\cite{Edelstein1996} and circadian rhythms~\cite{Leloup1999}, respectively, where the cost function measures the deviation from simulated data.
(Here, the derivatives in equation~\ref{eqn:hess} are taken in the logarithms of the biochemical parameters to reflect relative changes in parameter values.)
In all three cases the eigenvalues span more than $10^6$, indicating that the models are over one thousand times more sensitive to some directions than others; that the ellipses in figure~\ref{fig:banana}A are over one thousand times as long as they are wide.

Sloppiness is not restricted to biology, as illustrated by column (iv) in figure~\ref{fig:banana}B, which shows the eigenvalues for fitting parameters of a wave function for use in Quantum Monte-Carlo.
Nor is sloppiness restricted to very complex models~\cite{Waterfall2006}; column
(v) shows the spectrum for fitting the decay rates in a sum of 48 exponentials (appendix~\ref{sec:exp}).
Not all models are sloppy, however, column (vi) shows the eigenvalue spectra for
fitting a plane to a set of data (a typical form of multiple linear regression).
The eigenvalues all have roughly the same magnitude, indicating that all directions in parameter space are similarly sensitive, so the model is not sloppy. The presence of sloppiness in such a diverse range of nonlinear models (i---v) suggests that it is a universal feature of nonlinear multiparameter models.

\section{Prediction uncertainties}

Given that sloppiness is a common feature of complex biological models, and
nonlinear multiparameter models in general, we ask how sloppiness impacts making predictions.

\subsection{Constraining parameters}

The first step in making predictions from a model is constraining the model's parameters.
Because biological models are sloppy, predictions are generally much more efficiently constrained by collectively fitting model parameters than by directly measuring them~\cite{Gutenkunst2006}.
This can be understood from figure~\ref{fig:banana}A.
Fitting parameters to data naturally constrains the region of parameter sets consistent with that data (indicated by the dots) to lie along directions to which the model is insensitive.
Thus the parameter sets encompass relatively few model behaviors and predictions have small uncertainties.
Because the model is sloppy, predictions can have small uncertainties in spite of large regions of parameter uncertainty, as long as that parameter uncertainty is correlated along sloppy directions.

By contrast, direct parameter measurements yield uncorrelated parameter uncertainties.
For example, if $\theta_2$ were known less precisely than $\theta_1$, the region of acceptable parameter sets would be a vertical ellipse in figure~\ref{fig:banana}A.
We find that generally very few bare parameter directions are sloppy directions, so, unless the parameters have been measured very precisely, such an ellipse will encompass many behaviors and predictions uncertainties will be correspondingly large.

\subsection{Estimation algorithms}

After optimizing to find the best-fit set of parameters, prediction uncertainties can be calculated by accounting for model behavior over the region of parameter space that is consistent with the data.
Here we consider two approaches to calculate uncertainties: linearized covariance analysis (LCA) and Monte-Carlo analysis (MCA).

\subsubsection{Linear covariance analysis}
Linear covariance analysis involves two approximations: a quadratic expansion of the cost function about the best-fit parameters and a linear approximation of the model response to parameter changes.
The standard deviation, $\sigma_y$, of the prediction $y$ is then given by
\begin{equation}
\sigma_y^2 = \sum_{i,j} \left.\frac{\partial y}{\partial \theta_i} \, \left(H^{-1}\right)_{ij} \, \frac{\partial y}{\partial \theta_j}\right|_{\theta^*}.
\end{equation}
LCA is computationally inexpensive, particularly in problems where the cost function is a sum of squared residuals ($C\left(\theta\right) = \sum_k r_k\left(\theta\right)^2$).
Then the Hessian matrix can be approximated by the Fisher Information Matrix:
\begin{equation}
F_{ij}\left(\theta\right) = \sum_k \left. \frac{\partial r_k}{\partial \theta_i} \frac{\partial r_k}{\partial \theta_j} \right|_{\theta} \approx H_{ij}\left(\theta\right).
\end{equation}
This approximation is very useful because for many models (including those based on differential equations) first derivatives can be obtained semi-analytically, and obtaining stable finite-difference derivatives is difficult in sloppy problems.

\subsubsection{Monte-Carlo analysis}
Monte-Carlo analysis explicitly samples from the distribution of parameter sets consistent with the available data.
The posterior distribution of parameter sets given model $M$ and data $D$ is given by:
\begin{equation}
P\left(\theta|D;M\right) \propto \exp\left(-C\left(\theta\right)\right).\label{eqn:posterior}
\end{equation}
After an ensemble of parameter sets has been drawn from this distribution, the uncertainty on any quantity can be estimated simply by calculating it for each member of the parameter ensemble and noting the resulting variation.

Numerous algorithms exist to sample distributions like~(\ref{eqn:posterior}), but obtaining a well-converged ensemble can be computationally challenging.
After comparing predictions from LCA and MCA we will discuss specific algorithms and speed-ups for MCA.
Importantly, however, once obtained for a given model and set of constraining data the ensemble can be used for any number of predictions.

\subsubsection{Comparison}
\begin{figure}
\begin{center}
\includegraphics{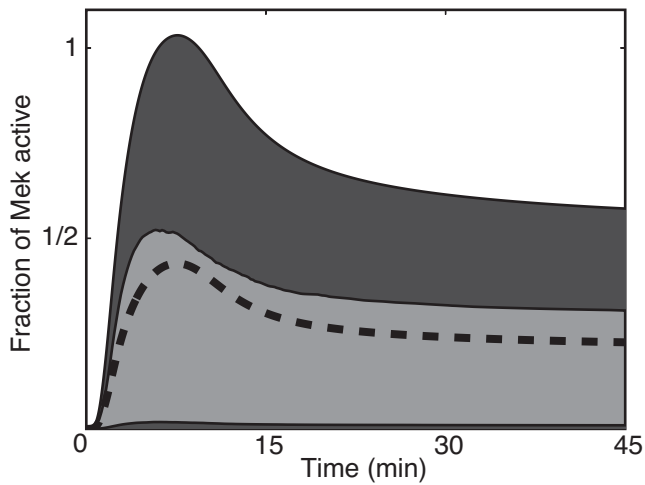}
\end{center}
\caption{Comparison between linearized covariance analysis (LCA) and Monte-Carlo analysis.
In dark grey is the one-standard-deviation uncertainty bound from LCA for the activity of Mek given NGF stimulation in our growth factor signaling model~\cite{Brown2004}. In light grey is the corresponding Monte-Carlo result. The dashed line is prediction from the best-fit set of parameters. Note how dramatically the LCA prediction overestimates the uncertainty.}\label{fig:method_comp}
\end{figure}

The approximations involved in LCA may introduce significant artifacts, as illustrated in figure~\ref{fig:method_comp}, which plots a particular prediction of our growth factor signaling model~\cite{Brown2004}.
LCA dramatically over-estimates the uncertainty in the prediction, particularly at early times, reducing the power of the prediction to test the model structure.
For example, an experimental result of 3/4 activity at 8 minutes lies outside the bounds of the well-converged MCA prediction and would suggest that some assumption in the model structure is probably incorrect.
The LCA prediction however, would not offer this insight.

In figure~\ref{fig:method_comp}, LCA overestimated the uncertainty, but in some cases LCA may also underestimate uncertainty.
Such underestimation may be even more damaging, as it may cause one to reject a model that MCA would reveal is actually consistent with the data.

The difference between LCA and MCA arises from nonlinearities in the cost surface and the model response.
In particular, the sloppy directions determined by the Hessian matrix are exactly those
directions that have small quadratic components in the cost.
Therefore in these directions higher order terms are responsible for constraining the behavior. 
We observe that prior information on parameter ranges can mimic the effect of these higher order terms, allowing the LCA to produce uncertainty estimates comparable to MCA.
Further, we are investigating the use of curvature measures of nonlinearity~\cite{Bates1980} to predict when the LCA will be inaccurate for sloppy models.

\subsection{Efficient Monte-Carlo for Sloppy Models}

Although sloppiness implies that many sets of parameters will be consistent with the data, those consistent sets remain a very small fraction of the entire parameter space.
Directly sampling from the posterior distribution (equation~\ref{eqn:posterior}) is thus infeasible.
We use Markov-Chain Monte-Carlo to sample the distribution, building our ensemble via a random walk through parameter space~\cite{Chib1995}.

For sloppy models the cost function is very stiff in some directions and very sloppy in others.
If we take steps at the scale of the stiff directions, exploring the sloppy directions will be very slow, but if the step size is too large moves in the stiff directions will cause very few steps to be accepted, also slowing convergence.
Thus it is vital to use \emph{importance sampling}, taking larger steps in some directions than others. 
We find that it is natural to scale steps using the square root of the regularized Fisher Information Matrix and that this gives reasonable acceptance probabilities.
Furthermore, we find it important to recalculate the FIM periodically to maintain a reasonable chance of accepting attempted moves.
This indicates that the cost surface is not only very narrow, but also substantially curved.

Building parameter ensembles in sloppy systems can also challenge algorithms that solve the model equations.
For example, the ensemble built for our differential equation model of growth factor signaling~\cite{Brown2004} explored regions of parameter space where the equations were very difficult to integrate, even for well-tested stiff integrators.
Resulting errors must be caught, and we found it useful to dynamically tighten integration tolerances.
Significantly for biochemical networks, Michaelis-Menten and Hill type equations
can become particularly difficult to numerically integrate when exploring
parameter space because the ``turning points'' in these curves can become
incredibly sharp and without sufficient guidance the integrator can overstep
these points and land on a different, unphysical solution to the equations.

\section{Standards for the computational systems biology community}

The Systems Biology Markup Language (SBML)~\cite{Hucka2003} has emerged as a community standard for the definition of models of biomolecular reaction networks.
By making our own code --- SloppyCell~\cite{SloppyCell} --- able to read and write SBML models, we have been able to apply our analyses to a number of published models in the literature, demonstrating the ubiquity of sloppiness~\cite{Gutenkunst2006}.
We found that this standard, while adequate for summarizing finished models, is inadequate for capturing the larger process of reverse engineering, in particular the derivation of model structure and parameter values from data.
In addition, since biological experimentation often involves creating topological variants (mutants) of a wild-type, biological modeling requires working with multiple related networks which collectively define a ``model''.
While SBML contains constructs for defining networks, it provides no data structures or operations for modifying or relating networks.
The biological reverse engineering community will --- at some stage --- need to address these deficiencies in current standards if community progress is to continue.

\section{Conclusions}

Model validation is a crucial part of reverse-engineering biological networks, and making predictions is one of the most important parts of validation.
To usefully test model structures, predictions must be accompanied by uncertainty estimates that account for the underlying parameter uncertainty.
Linearized covariance analysis may introduce important artifacts, so when possible Monte-Carlo analysis should be preferred, and such analysis has specific numerical challenges that need to be overcome.
As the community and associated standards develop we anticipate that techniques to make and test predictions will prove increasingly important.

\appendix
\section{Fitting exponentials}\label{sec:exp}
The fitting of a sum of exponentials is a classic problem~\cite{lanczos}, and, as seen in figure~\ref{fig:banana}B it is sloppy.
Fitting exponentials has proven a very useful test problem both for analysis and
for algorithm development.
It is physically motivated, but computationally convenient; the cost and Hessian matrix are simple functions of the true decay rates and initial amounts.
This facilitates generation of high quality statistics to investigate questions concerning sloppiness, such as the relationship between parameter degeneracy and the spacings between eigenvalues~\cite{Waterfall2006}. It also makes fitting exponentials a useful test case for comparing and refining optimization algorithms.

\bibliography{sloppy}
\end{document}